\documentclass[sigconf,nonacm]{acmart}
\usepackage{amsfonts}
\usepackage{amsmath}
\usepackage{graphicx}
\usepackage{subcaption}
\usepackage[braket,qm]{qcircuit}
\usepackage{braket}
\usepackage{tikz}

\newcommand{\cO}{\mathcal{O}}

\newcommand{\cS}{\mathcal{S}}
\newcommand{\cT}{\mathcal{T}}

\begin{document}

\title{Exploiting recursive structures for the design of novel quantum primitives}

\author{Ning Bao}
\email{ningbao75@gmail.com}
\affiliation{%
    \institution{Brookhaven National Laboratory}
    \city{Upton, NY}
    \country{USA}
}
\affiliation{%
    \institution{Northeastern University}
    \city{Boston, MA}
    \country{USA}
}

\author{Gün Süer}
\email{suer.g@northeastern.edu}
\affiliation{%
  \institution{Northeastern University}
  \city{Boston, MA}
  \country{USA}
}

\begin{abstract}
The advent of fault-tolerant quantum computers marks a significant milestone, yet the development of practical quantum algorithms remains a critical challenge. Effective quantum algorithms are essential for leveraging the power of quantum computers, and their design is often non-intuitive. This paper addresses the issue of generating novel quantum primitives by focusing on recursive circuits. We explore the recursive circuit structures prevalent in existing quantum algorithms and demonstrate how these structures can be exploited to design new, potentially advantageous quantum algorithms. We base our discussion on the quantum Fourier transform (QFT), which is a primitive that is widely used in quantum algorithms. We show that the recursive structure in well-established fast classical transforms forms a fruitful bridge with quantum algorithms, enabling the design of novel quantum primitives and the discovery of new discrete numerical transforms. The discussion is split into two complementary parts, the forward and the reverse direction, in which existing classical transforms are implemented using polynomial-time quantum circuits and recursive circuits are used to find novel non-sparse classical transforms with guaranteed quantum speedup, respectively. We comment on the potential impact on quantum algorithms, numerical analysis, and signal processing.

\end{abstract}

\maketitle

\section{Introduction}

Quantum algorithms provide computational speedup over their classical counterparts by encoding information in qubits, leveraging superposition and entanglement, and then processing them using often-unitary quantum circuits \cite{kitaev2002classical}. Some of the most notable ones include Shor's algorithm \cite{Shor_1997}, quantum phase estimation \cite{kitaev1995quantummeasurementsabelianstabilizer}, Grover search \cite{grover1996fastquantummechanicalalgorithm}, and the Harrow–Hassidim–Lloyd (HHL) algorithm \cite{Harrow_2009}. Even though this list can be extended, the majority of the existing algorithms are derived from only a handful of quantum primitives, such as the quantum Fourier transform (QFT) and the Grover search iterator. This situation promotes the idea that the range of applicability of quantum algorithms is bottlenecked by the number of distinct quantum primitives. This is not surprising, since the design of quantum algorithms that implement desired unitary transformations is non-intuitive, and at present there is no straightforward way to create efficient quantum circuits for interesting problems.

\begin{table}
    \begin{tikzpicture}
    
    \node[draw, minimum height=1cm, minimum width=2cm, align=center] (A) at (0,1) {Fast\\classical\\transforms};
    \node[draw, minimum height=1cm, minimum width=2cm, align=center] (C) at (0,-1) {Numerical\\classical\\transforms};
    \node[draw, minimum height=1cm, minimum width=2cm, align=center] (B) at (4,1) {Novel\\quantum\\primitives};
    \node[draw, minimum height=1cm, minimum width=2cm, align=center] (D) at (4,-1) {Recursive\\quantum\\circuits};

    \draw[->, line width=.5mm] (A.east) to[out=45,in=135] (B.west);
    \draw[->, line width=.5mm] (D.west) to[out=225,in=315] (C.east);

    \node[align=center] at (2,2) {Forward\\direction};
    \node[align=center] at (2,-2) {Reverse\\direction};
    
    \end{tikzpicture}
    \caption{Diagrammatic representation of the two directions covered in this paper. Starting from fast classical transforms we can implement them using quantum circuits to find novel quantum primitives. In the other direction, starting with a recursive quantum circuit, we arrive at novel classical transforms with guaranteed speedup.}
    \label{tab:my_label}
\end{table}

We focus on automating the design of quantum primitives by understanding the common structures they possess. A fruitful one in this direction turns out to be recursivity, which is already present in the QFT. It is well known that the recursive nature of the discrete Fourier transform is what enables the formulation of fast classical transforms of the radix-2 type \cite{cooley1965algorithm}. This property viewed in matrix form provides a straightforward way of understanding how the QFT circuit is designed, and streamlines how other fast classical transforms can be implemented via a quantum circuit. Approaching this connection from the other direction, we show that any quantum circuit that is recursive yields an exponential speedup over any fast classical transform trying to implement it's unitary. Hence just by changing the building blocks of the recursive circuit, one arrives at multitudes of different classical transform with guaranteed quantum speedups. In this sense, recursivity is the main theme that ties fast classical transforms and quantum algorithms, and utilizing it in a brute-force manner has the potential for the discovery of both novel quantum primitives and classical transforms. These directions are summarized in Table \ref{tab:my_label}. Therefore, this research direction is expected to have an impact in quantum algorithms, numerical analysis, and signal processing.

The organization of this paper is as follows. In Section \ref{sec:forward}, we start with a gentle reminder of the classical and quantum Fourier transform. We briefly show how the quantum circuit implementing the QFT is derived from its product form, and comment on how the recursive structure of the circuit leads to a gate complexity that is polynomial in the number of qubits. We compare the recursion relation in the circuit with another that can be found in the matrix form of the discrete transform. We emphasize that this constitutes the backbone of the fast Fourier transform (FFT) \cite{cooley1965algorithm}. In Section \ref{sec:reverse}, we approach the problem in the reverse direction, starting with a recursive quantum circuit that is generated by arbitrary building blocks. We show that regardless of the chosen building blocks, the circuit always has gate complexity $\cO(n^2)$, and the unitary matrix that it implements is non-sparse. The lower bound for matrix-vector multiplication is given by $\cO(n 2^n)$, hence any new numerical transform that we discover using the recursive quantum circuit ansatz is guaranteed to have a quantum speedup. We also rule out any efficient classical implementations due to non-sparsity. We conclude the paper with Section \ref{sec:discussions} by discussing applications to quantum algorithms, numerical analysis, signal processing, and future research directions.

\section{Forward direction}
\label{sec:forward}

With these considerations in mind, we resort to a brute force approach for generating potentially useful quantum algorithms that have strong speedup probability. We draw inspiration from the implementation of the QFT, which we will review here for pedagogical clarity. First, recall the discrete Fourier transform \cite{oppenheim1975digital}, a ubiquitous classical transform in physics and engineering that represents a digital signal $\{x_k\}$ for $k=0,1,\dots, N-1$ in frequency domain $\{X_k\}$ given by
\begin{equation}
    X_k = \sum_{m=0}^{N-1} e^{-i\frac{2\pi}{N}km} x_m.
\end{equation}
The discrete transform can equivalently be described as a matrix $F$, with $F_{km}=e^{-i\frac{2\pi}{N}km}$ that acts like $\vec{X} = F \vec{x}$.

By encoding the signal as amplitudes of a quantum state, a signal of size $N=2^n$ can be represented using an $n$-qubit quantum state \cite{nielsen2000quantum}, and the QFT acts as
\begin{equation}
    \ket{k} \to \frac{1}{2^{n/2}}\sum_{m=0}^{2^n-1} e^{i\frac{2\pi}{N}km} \ket{m}.
\end{equation}
By representing the basis vectors using $n$-bit binary numbers, i.e. $k = k_1 2^{n-1} + k_2 2^{n-2} + \cdots + k_n 2^0$, the QFT can be written in a product form of single qubit states
\begin{equation}
    \ket{k_1,\cdots,k_n} \to \frac{\left(\ket{0}+e^{2\pi i 0.k_1}\ket{1}\right)\cdots\left(\ket{0}+e^{2\pi i 0.k_1\dots k_n} \ket{1}\right)}{2^{n/2}}
\end{equation}
where we used the binary fraction notation $0.k_l k_{l+1} \dots k_m = k_l/2 + k_{l+1}/2^2 + \cdots + k_m/2^{m-l+1}$. Written in the product form, it is easy to see each gate that must be applied to individual qubits. For example, the circuit implementation for the 4-qubit QFT is given by Table \ref{circuit:qft},
\begin{table}
\[
\Qcircuit @C=1em @R=1em {
    & \gate{H} & \gate{R_2} & \gate{R_2} & \gate{R_3} & \qw & \qw & \qw & \qw & \qw & \qw & \qw  \\
    & \qw  & \ctrl{-1} & \qw & \qw & \gate{H} & \gate{R_2} & \gate{R_3} & \qw & \qw & \qw & \qw \\
    & \qw  & \qw  & \ctrl{-2} & \qw & \qw & \ctrl{-1} & \qw & \gate{H} & \gate{R_2} & \qw & \qw \\
    & \qw & \qw & \qw & \ctrl{-3} & \qw & \qw & \ctrl{-2} & \qw & \ctrl{-1} & \gate{H} & \qw
}
\]
\caption{Circuit diagram for the 4-qubit quantum Fourier transform.}
\label{circuit:qft}
\end{table}
where $H$ is the Hadamard gate and the phase gate is given by Equation \eqref{eq:phase_gate}
\begin{equation}
\label{eq:phase_gate}
    R_k =
    \begin{pmatrix}
        1 & 0\\
        0 & e^{2\pi i / 2^k}
    \end{pmatrix}
\end{equation}

Due to the recursive structure of the quantum circuit, adding an additional qubit changes the gate complexity by $n$, therefore the QFT acting on $n$-qubits has gate complexity $\cO(n^2)$. Since we can encode a signal of size $N=2^n$ using $n$-qubits, the time complexity in terms of the signal size is given by $\cO(\log^2 N)$.

Let us compare the time complexity of the QFT with the FFT. The fast Fourier transform \cite{cooley1965algorithm} uses a divide and conquer method and the periodicity of the Fourier transform to relate the calculation of the $N$-point Fourier transform to the $\frac{N}{2}$-point Fourier transform. Splitting the Fourier transform to odd and even parts we obtain
\begin{align}
\label{eq:radix-2}
    X_k 
    &= \sum_{m=0}^{N/2-1} e^{-\frac{2\pi}{N}i (2m) k} x_{2m} + \sum_{m=0}^{N/2-1} e^{-\frac{2\pi}{N}i (2m+1) k} x_{2m+1} \nonumber\\
    &= \sum_{m=0}^{N/2-1} e^{-\frac{2\pi}{N/2}i m k} x_{2m} + e^{-\frac{2\pi}{N}i k} \sum_{m=0}^{N/2-1} e^{-\frac{2\pi}{N/2}i m k} x_{2m+1} \nonumber\\
    &=: E_k + e^{-\frac{2\pi}{N}i k} O_k
\end{align}
where we defined the Fourier transform over the even and odd indices as $E_k$ and $O_k$ respectively. Due to the periodicity of the complex exponential, $X_{k+\frac{N}{2}}$ is given by another linear combination of the Fourier transform over the even and odd indices
\begin{equation}
    X_{k+\frac{N}{2}} = E_k - e^{-\frac{2\pi}{N}i k} O_k
\end{equation}
Therefore the $N$-point Fourier transform can be computed by repeatedly dividing the signal in half. Since each sum can be divided in half $\log_2 N$ number of times, and each sum has $N$ additions, the total time complexity is given by $\cO(N\log N)$.

One can also utilize the radix-2 type recursion relation in the matrix form \cite{yuen_website_qft}, which will be useful in constructing the quantum circuit implementing a classical transform. The $N\times N$ Fourier transform can be written in terms of the $\frac{N}{2}\times \frac{N}{2}$ Fourier transform as
\begin{equation}
    \label{eq:matrix-recursion}
    F_N = \frac{1}{\sqrt{2}}
    \begin{pmatrix}
        F_{\frac{N}{2}} & A_{\frac{N}{2}} F_{\frac{N}{2}}\\
        F_{\frac{N}{2}} & -A_{\frac{N}{2}} F_{\frac{N}{2}}
    \end{pmatrix}
\end{equation}
where the matrix $A_{\frac{N}{2}}$ contains the twiddle factors along its diagonal
\begin{equation}
    A_{\frac{N}{2}} =
    \begin{pmatrix}
        1 & & & & \\
          & e^{-\frac{2\pi}{N}i} & & & \\
          & & e^{-2\frac{2\pi}{N}i} & & \\
          & & & \ddots & \\
          & & & & e^{-(\frac{N}{2}-1)\frac{2\pi}{N}i}
    \end{pmatrix}
\end{equation}
and the smallest Fourier transform is given by
\begin{equation}
    F_2 = \frac{1}{\sqrt{2}}
    \begin{pmatrix}
        1 & 1\\
        1 & -1
    \end{pmatrix}
\end{equation}
From the even odd decomposition of the radix-2 algorithm \eqref{eq:radix-2}, it's easy to see that $A_{\frac{N}{2}}$ is given by the tensor product of phase matrices \eqref{eq:phase_gate} we have previously defined
\begin{equation}
    A_{\frac{N}{2}} = R_2 \otimes R_3 \otimes \cdots \otimes R_N
\end{equation}
Therefore, any fast classical transform can be implemented as a quantum circuit by finding the correct $A$-matrix, and repeatedly applying it to decreasing number of qubits, yielding novel quantum algorithms.

\section{Reverse direction}
\label{sec:reverse}

We have reviewed the Fourier transform in the previous section, and we have seen that while the FFT saturates the lower bound $\cO(N\log N)$ on matrix-vector multiplication \cite{10.1145/509907.509932}, the QFT is still exponentially faster than the FFT with gate complexity $\cO(\log^2 N)$. Upon inspection we understand that this is exponential speedup is not specific to the QFT, but any circuit that has the ladder structure\footnote{With ladder circuits we refer to quantum circuits that require a linear overhead to implement an $n$-qubit transform using the $(n-1)$-qubit transform.} has a log-squared gate complexity in the number of qubits.

Viewed in the reverse direction of the previous section, we can take the recursive circuit ansatz, given in Table \ref{tab:circuit-ansatz}, and change the single-qubit and two-qubit gates that are used in $V_n$. The resulting circuit has $\cO(\log^2 N)$ gate complexity, while its unitary can only be implemented classically at least at time $\cO(N\log N)$. Therefore the novel classical transform that is defined with an arbitrary choice of $V_n$ has a guaranteed exponential speedup.\footnote{The unitary matrices implemented by ladder circuits are non-sparse. Therefore they don't have efficient classical simulations that could violate the $\cO(N\log N)$ lower-bound.}

Generalizing the ladder structure of the QFT, a generic recursive circuit can be written in the following form
\begin{equation}
    U_n = \prod_{i=0}^{n-1} I^{\otimes i} \otimes V_{n-i}
\end{equation}
where $V_i$ is the overhead linear in the number of qubits required to implement the recursive circuit for $n$-qubits using the quantum circuit for $(n-1)$-qubits. This is illustrated in Table \ref{tab:circuit-ansatz}. Therefore $V_i$ and $U_j$ satisfy the following recursion relation
\begin{equation}
    U_n = (I \otimes U_{n-1}) V_n 
\end{equation}
Therefore a recursive circuit can be completely determined by the choice of 1-qubit and 2-qubit gates that we use in $\{V_i\}$
\begin{equation}
    V_{n-i} = \cS_i \prod_{j=1}^{i-1} \cT_{(i,j)}
\end{equation}
where $\cS_i$ is a 1-qubit gate that acts on qubit-$i$, and $\cT_{(i,j)}$ is a traversal 2-qubit gate that acts on qubit-$i$, controlled by qubit-$j$.

Numerical transforms implemented using these techniques are generally non-sparse, due to the recursive structure of the circuit. At each iteration of the $\{V_i\}$'s, a 2-qubit gate is applied which acts on the $i$-th qubit, controlled by the rest of the qubits, in addition to the preceding 1-qubit gate. This structure promotes the coupling between different qubits in the system. Provided that the single qubit unitary is non-sparse, the coupling provided by the 2-qubit gates populate the entire matrix.\footnote{A common choice is the Hadamard operator for the single qubit gate.} Therefore it is highly unlikely for these circuits to implement sparse unitary matrices, which renders any fast classical implementation improbable. Therefore the reverse direction with the choice of correct gates provide true quantum speedup. Different gate choices that we studied numerically in this work have been listed in Table \ref{tab:sparsity}. The non-sparsity realized is exact.

\begin{table}
\[
\Qcircuit @C=1em @R=1em {
    & \qw & \multigate{4}{V_n} & \qw & \qw & \qw & \qw & \qw\\
    & \qw & \ghost{V_n}  & \multigate{3}{V_{n-1}}& \qw & \qw & \qw & \qw \\
    & \qw & \ghost{V_n} & \ghost{V_{n-1}} & \ddots & & & & &\\
    & \qw & \ghost{V_n}  & \ghost{V_{n-1}}  & \qw & \multigate{1}{V_2} & \qw & \qw \\
    & \qw & \ghost{V_n} & \ghost{V_{n-1}} & \qw & \ghost{V_2} & \gate{V_1} & \qw
}
\]
\caption{A generic recursive circuit. To get the $n$-qubit circuit from the $(n-1)$-qubit circuit, one needs to act with $V_n$, which has gate complexity linear in the number of qubits.}
\label{tab:circuit-ansatz}
\end{table}

By exploiting the recursive structure of a family of quantum circuits we have exchanged the problem of designing efficient quantum circuits for unitary matrices with interesting applications with the problem of finding quantum circuits that potentially solve interesting problems from a vast array of circuits which have guaranteed speed-up over their classical counterpart, that we have generated using brute force methods, inspired by the recursive structure in the QFT.

\section{Numerical results}
We provide numerical tests for reverse direction by considering recursive circuits that yield novel discrete classical transforms. By changing the 1-qubit and 2-qubit gates, the recursive circuits implement unitary matrices that are potentially new with uncapitalized applications. Due to space constraints, these transforms are represented by a heat-map of it's matrix entries.

\begin{figure}[h!]
    \centering
    \begin{subfigure}[b]{0.45\columnwidth}
        \centering
        \includegraphics[width=\linewidth]{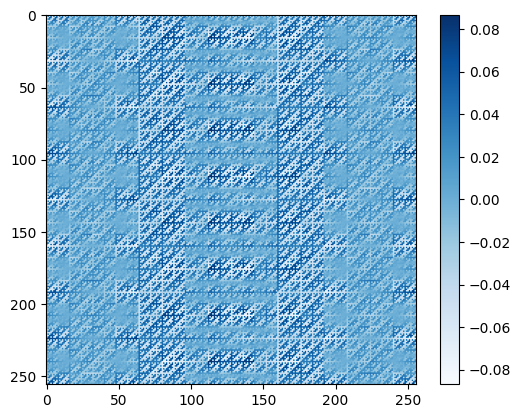}
        \caption{IsingXX with H}
    \end{subfigure}
    \hfill
    \begin{subfigure}[b]{0.45\columnwidth}
        \centering
        \includegraphics[width=\linewidth]{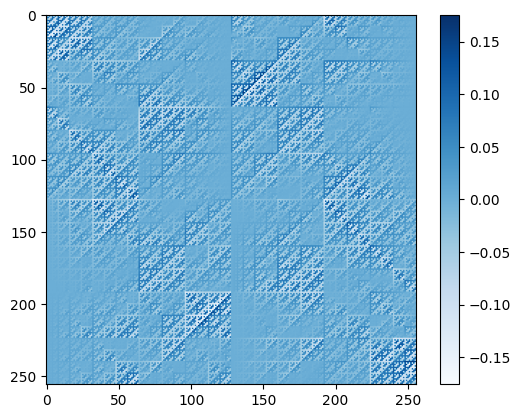}
        \caption{IsingXY with H}
    \end{subfigure}
        
    \begin{subfigure}[b]{0.45\columnwidth}
        \centering
        \includegraphics[width=\linewidth]{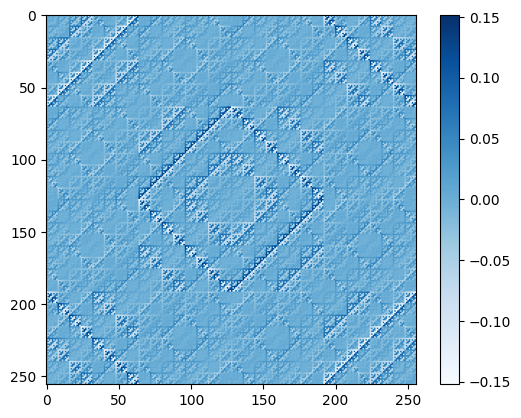}
        \caption{IsingYY with H}
    \end{subfigure}
    \hfill
    \begin{subfigure}[b]{0.45\columnwidth}
        \centering
        \includegraphics[width=\linewidth]{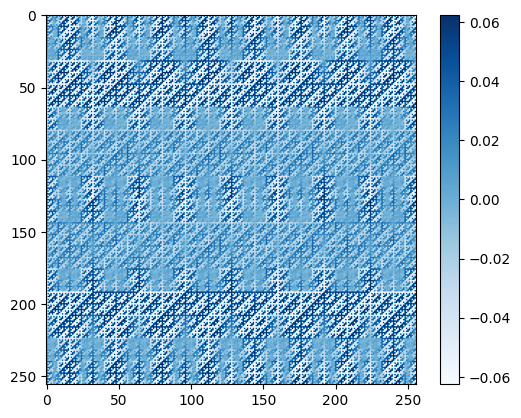}
        \caption{IsingZZ with H}
    \end{subfigure}
    
    \caption{Classical matrices for a fixed number of qubits $\log_2 N = 8$ for different choice of generating gates. Only real part of the unitary matrices are shown.}
    \label{fig:3}
\end{figure}

\begin{figure}[h!]
    \centering
    \begin{subfigure}[b]{0.45\columnwidth}
        \centering
        \includegraphics[width=\linewidth]{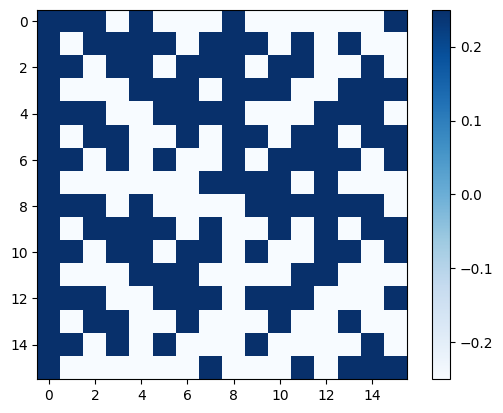}
        \caption{$\log_2 N = 4$}
    \end{subfigure}
    \hfill
    \begin{subfigure}[b]{0.45\columnwidth}
        \centering
        \includegraphics[width=\linewidth]{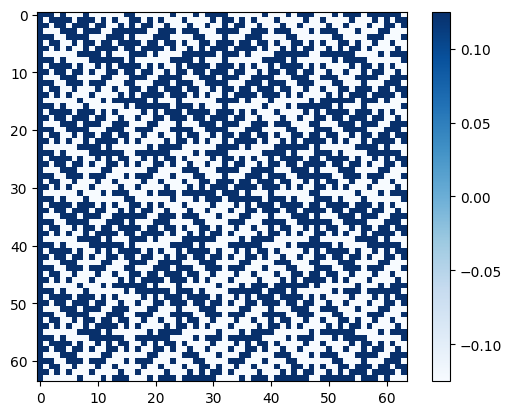}
        \caption{$\log_2 N = 6$}
    \end{subfigure}
        
    \begin{subfigure}[b]{0.45\columnwidth}
        \centering
        \includegraphics[width=\linewidth]{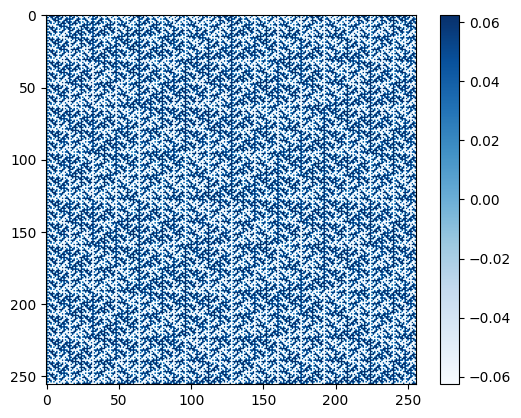}
        \caption{$\log_2 N = 8$}
    \end{subfigure}
    \hfill
    \begin{subfigure}[b]{0.45\columnwidth}
        \centering
        \includegraphics[width=\linewidth]{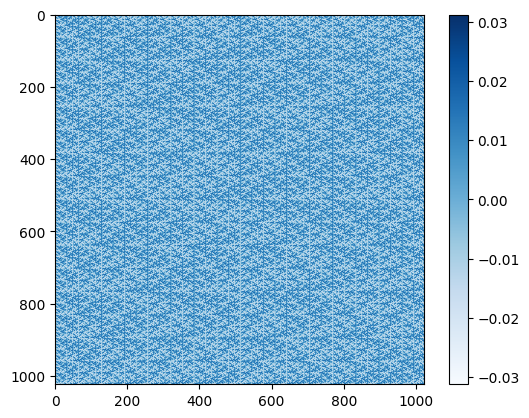}
        \caption{$\log_2 N = 10$}
    \end{subfigure}
    
    \caption{Classical matrices corresponding to recursive circuit generated by Hadamard and CNOT gates, for increasing number of qubits $\log_2 N = 4, 6, 8, 10$. Only real part of the unitary matrices are shown.}
    \label{fig:1}
\end{figure}

\begin{figure}[h!]
    \centering
    \begin{subfigure}[b]{0.45\columnwidth}
        \centering
        \includegraphics[width=\linewidth]{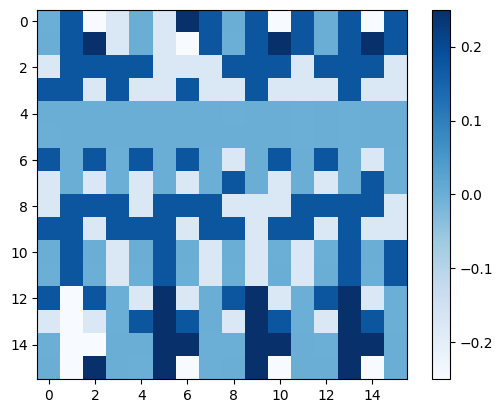}
        \caption{$\log_2 N = 4$}
    \end{subfigure}
    \hfill
    \begin{subfigure}[b]{0.45\columnwidth}
        \centering
        \includegraphics[width=\linewidth]{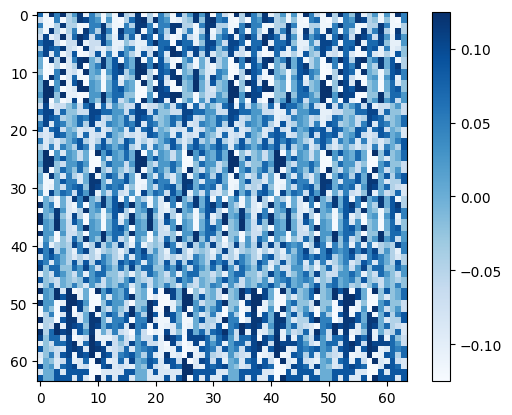}
        \caption{$\log_2 N = 6$}
    \end{subfigure}
        
    \begin{subfigure}[b]{0.45\columnwidth}
        \centering
        \includegraphics[width=\linewidth]{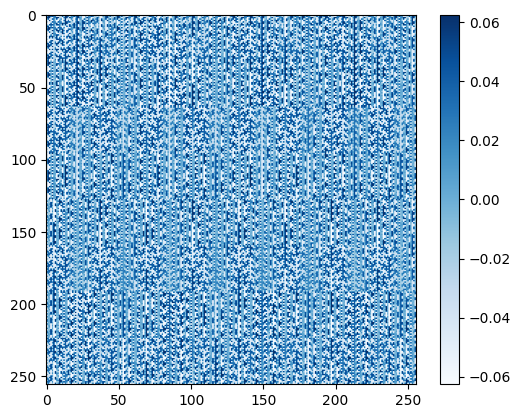}
        \caption{$\log_2 N = 8$}
    \end{subfigure}
    \hfill
    \begin{subfigure}[b]{0.45\columnwidth}
        \centering
        \includegraphics[width=\linewidth]{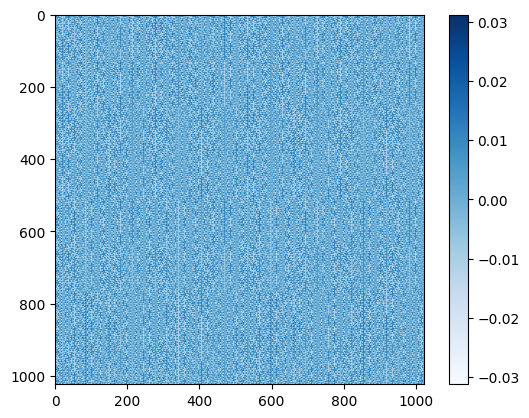}
        \caption{$\log_2 N = 10$}
    \end{subfigure}
    
    \caption{Classical matrices corresponding to recursive circuit generated by Hadamard, CNOT, and controlled phase shift gates, for increasing number of qubits $\log_2 N = 4, 6, 8, 10$. Only real part of the unitary matrices are shown.}
    \label{fig:2}
\end{figure}

For initial tests we consider single-qubit Hadamard gates and two-qubit Ising coupling gates. The discrete transforms generated by these gate sets are shown in Figure \ref{fig:3}, for fixed number of qubits. We also consider how the gradual increase in qubit number changes the pattern in the classical matrix. To that end we consider a circuit generated by Hadamard and CNOT gates, and record the consequence of increasing the number of qubits. The change in pattern as we increase the number of qubits highlight the recursive structure present, as shown in Figure \ref{fig:1}. We perform a similar demonstration for a more complicated set of gates in Figure \ref{fig:2}.

The emerging structure is demonstrated by increasing the number of qubits in Figures \ref{fig:1} and \ref{fig:2}. The transforms generated by using the Ising coupling gates are represented in Figure \ref{fig:3} for fixed number of qubits.

\begin{table}[h!]
\centering
\begin{tabular}{ |p{1.0cm}|p{3.5cm}||p{1.0cm}|p{1.0cm}| }
 \hline
 \multicolumn{4}{|c|}{Choice of different generating gates} \\
 \hline
 $\cS(i)$ & $\cT_{(i,j)}$ & $\log_2 N$ & Non-sparse\\
 \hline
 $T$   &  $CX_{(i,j)}$ & 4 & \checkmark\\
    &   & 8 & \checkmark\\
    &   & 10 & \checkmark\\
 \hline
 $H$   &  $CX_{(i,j)} CP_{(i,j)}(2\pi/2^j)$ & 4 & \checkmark\\
    &   & 8 & \checkmark\\
    &   & 10 & \checkmark\\
 \hline
 $HX$   &  $CP_{(i,j)}(2\pi/2^j)$ & 4 & \checkmark\\
    &   & 8 & \checkmark\\
    &   & 10 & \checkmark\\
 \hline
 $H$   &  $\exp(i\frac{\pi}{2^j}(X_i\otimes X_j + Z_i\otimes Z_j))$ & 4 & \checkmark\\
    &   & 8 & \checkmark\\
    &   & 10 & \checkmark\\
 \hline
 $T$   &  $\exp(-i\frac{\pi}{2^j} Z_i \otimes Z_j)$ & 4 & \checkmark\\
    &   & 8 & \checkmark\\
    &   & 10 & \checkmark\\
 \hline
\hline
\end{tabular}
\caption{Different gate choices used in numerical simulations. We tested these choice over different number of qubits, and found that they result in non-sparse unitary matrices.}
\label{tab:sparsity}
\end{table}

\section{Discussion}
\label{sec:discussions}

In this paper, we started with a pedagogical discussion of the QFT transform, and underlined the key recursive properties that enables its fast classical implementation as well as the construction of its quantum circuit. Inspecting the matrix level recursion relation that the DFT possesses, we showed how a radix-2 type fast classical transform can be implemented as a quantum circuit, that is exponentially faster than its classical counterpart, and how this leads to the discovery of novel quantum primitives. In a complementary direction, we used the ladder circuit ansatz to generate novel fast classical transforms with guaranteed speedups, by modifying the 1-qubit and 2-qubit gates used in each iteration.

For future directions, we hope to report on concrete implementations of existing fast classical transforms as quantum algorithms, as outlined in \ref{sec:forward}. Our work on the quantum Hermite transform is soon to appear \cite{bao_hermite}. We will be following this work with other fast classical transforms, and understanding their potential applications.

Another important question is of interpretability for the reverse direction. As it stands, the method yields novel non-sparse discrete transforms from recursive quantum circuits with proven quantum speedup, but their applications are unknown. Therefore we are faced with the problem of interpretability. We hope to study the effect of the novel classical transforms that were discussed in paper in analyzing discrete time signals, as well as their use in post processing images. Machine learning methods can also be used in characterizing these discrete transforms, and extrapolating their analytical forms if they exist. One can also introduce order to the novel classical transforms by requiring they minimize or maximize certain matrix norms, e.g. $L_1$-norm, so we would have to find the optimal gate sets such that the recursive circuit generated using them conform to some biases regarding classical matrix norm.

\section*{Acknowledgments}
We would like to thank Stephen Jordan and Troy Sewell for useful discussions. N.B. is supported by the DOE Office of Science-ASCR, in particular under the grant Novel Quantum Algorithms from Fast Classical Transforms. G. S. is supported by N.B.'s startup grant at Northeastern University.
\bibliographystyle{plain}
\bibliography{ref.bib}

\end{document}